# Recent progress towards acoustically mediated carrier injection into individual nanostructures for single photon generation

Stefan Völk[a], Florian J. R. Schülein[a], Florian Knall[a], Achim Wixforth[a], Hubert. J. Krenner*[a]

[a]Lehrstuhl für Experimentalphysik 1, Universität Augsburg, Universitätsstr. 1, 86159 Augsburg, Germany

*hubert.krenner@physik.uni-augsburg.de

Arne Laucht[b], Jonathan J. Finley[b]

[b]Walter Schottky Institut, Technische Universität München, Am Coulombwall 3, 85748 Garching, Germany

Juha Riikonen[c], Marco Mattila[c], Markku Sopanen[c], Harri Lipsanen[c]

[c]Department of Micro and Nanosciences, Micronova, Helsinki University of Technology, P.O. Box 3500, FIN-02015 TKK, Finland

Jun He[d], Tuan A. Truong[d], Hyochul Kim[e], Pierre M. Petroff[d,f]

[d]Materials Department; [e]Physics Department; [f]Department of Electrical and Computer Engineering
University of California, Santa Barbara CA 93106, United States

**ABSTRACT**

We report on recent progress towards single photon sources based on quantum dot and quantum post nanostructures which are manipulated using surface acoustic waves. For this concept acoustic charge conveyance in a quantum well is used to spatially separate electron and hole pairs and transport these in the plane of the quantum well. When conveyed to the location of a quantum dot or quantum post these carriers are sequentially captured into the confined levels. Their radiative decays gives rise to the emission of a train of single photons. Three different approaches using (i) strain-induced and (ii) self-assembled quantum dots, and (iii) self-assembled quantum posts are discussed and their application potential is discussed. First devices and initial experiments towards the realization of such an acoustically driven single photon source are presented and remote acoustically triggered injection into few individual emitters is demonstrated.

## 1. INTRODUCTION

Since the first demonstration of single photon emission from semiconductor quantum dots (QD) by Michler et al. in 2000 [1] these nanostructures have proven to be extremely suitable candidates for the implementation of quantum cryptography and quantum information processing schemes [2]. In particular, the ability to tune their emission wavelength from the visible to the telecom spectral range offers the potential for applications both in free-space and in fiber based realizations of a single photon source (SPS). The combination of QD emitters and photonic band gap materials and high quality optical cavities allow for an improved single photon extraction and a reduction of the exciton radiative lifetime [3-6]. Since the single photon generation was initially realized from the emission of a single exciton $1X^0$ formed by one electron and one hole, the efficiency could be substantially improved by avoiding optically inactive, "dark" excitons. This is because for the neutral exciton ($1X^0$) the spin of the electron and hole have to add up to a total angular momentum of ±1 to be optically active. Thus, adding an extra electron (hole) creates optically active, "bright" negatively $1X^-$ (positively $1X^+$) charged excitons. These approaches were successfully combined and yielded high single photon emission rates exceeding tens of MHz [7].

While quantum cryptography schemes can be implemented with reasonable fidelity even by SPS [8] employing attenuated laser beams, for advanced protocols such as quantum teleportation, a pure SPS is no longer sufficient. The required advanced schemes ask for an additional degree of freedom only accessible for two particle quantum states, namely entanglement [9]. Exploiting entanglement provides the potential to realize protocols with significantly improved security which can neither be obtained by simply using more bits for the encoding in standard cryptography nor by SPS based quantum cryptography.

For self-assembled quantum dots, a proof of principle of photon pair emission being entangled in the polarization domain has recently been made. It has been achieved using the so-called biexciton-exciton emission cascade where a QD occupied by two electrons plus two holes (biexciton) first decays into an intermediate one electron plus one hole (exciton) which finally decays and emits a second photon leaving behind an empty QD. Due to the optical selection rules and after preparing the QD by e.g. growth optimization or external fields to compensate for structural assymmetries, this pair of photons is entangled via their polarizations [10-12].

For any practical use, a deterministic generation of such initial biexciton state has to be ensured. This is particularly challenging since the generation via optical and electrical pumping typically leads to a significant occupation probability of different configurations (e.g. charged excitons with different numbers of electrons and holes). In the proposals by Imamoglu et al. [13] and Benson et al. [14], it was shown that in a *p-i-n* light emitting diode (LED) with precisely engineered tunnel barriers the number of electrons and holes in the dot could be tuned by the gate voltage and a sequential injection of the two carrier species occurs. The underlying effect is Coulomb blockade where injection of carriers from the contact becomes blocked via the electrostatic repulsion of the carriers already residing in the QD. The realization of this idea turned out to be extremely challenging: in particular the sequential injection of two electrons followed by two holes or vice versa. Despite much effort put into this idea the resulting "turnstile" could only be demonstrated at milli-Kelvin temperatures and not using QDs as emitters [15]. The crucial design parameter for such structures is clearly in the precise design of the barriers to achieve the sequential injection. If this is not the case, the fraction of non-suitable exciton species increases strongly or even dominates over the biexciton generation.

In this paper, we report on recent advances in the realization of a SPS employing an inherently different injection scheme based on surface acoustic waves (SAW). This approach is based on the Type-II band modulation induced by the SAW which spatially separates electrons and holes, hence suppressing their radiative decay. Due to the propagation of the SAW, these long-lived carriers can be transported over macroscopic distances in the plane of a quantum well giving rise to a charge conveyance effect [16]. We investigate the potential to implement this scheme for different QD systems and demonstrate the remote injection of electrons and holes into few and individual QD nanostructures. Furthermore, we discuss the ability to control the charge state and number of carriers in the QD and, finally, show directions to purely electrically operate such a SAW-driven SPS.

## 2. IMPLEMENTATION OF A SAW-DRIVEN SINGLE PHOTON SOURCE USING DIFFERENT QD-SYSTEMS

In piezoelectric materials, the mechanical deformation of a crystal lattice induced by a SAW gives rise to large electric fields with amplitudes exceeding 1 kV/cm both along (longitudinal) the propagation direction and normal to the surface of the material. In piezoelectric semiconductors such GaAs, AlAs and InAs the longitudinal component gives rise to a Type-II band edge modulation for which the total bandgap energy remains approximately constant and potential minima for electrons and holes are induced along the SAW propagation direction. As shown in the center part of Fig. 1 (a) these minima are separated by half of the SAW's wavelength ($\lambda_{SAW}/2$). If the carriers are additionally confined in the normal direction which for example can be realized by a Quantum Well (QW) heterostructure, these dynamically generated pockets of electrons and holes propagate with the SAW and give rise to a charge conveyance effect [16]. This effect can be used to transport carriers almost loss-free over macroscopic distances. The SAW induced charge conveyance principle was proposed to deterministically inject carriers into the energetically deeper levels of QDs "coupled" to the QW [17,18] which in turn emits single photons during each SAW cycle i.e. after injection of both carrier species. Such a SAW-driven SPS is depicted schematically in Fig. 1 (a). It consists of four major parts:

i. An interdigital transducer (IDT) to excite a SAW which propagates across a semiconductor heterostructure containing a QW.
ii. A generation area close to the IDT at which electrons and holes are created e.g. by optical excitation.
iii. A transport region for conveyance of spatially separated electrons and holes by the SAW.
iv. A QD nanostructure into which electrons and holes are sequentially injected. Thus, excitons are formed which radiatively recombine and emit single photons.

When compared to conventional injection schemes this SAW-based method offers inherently a sequential injection of electrons and holes. For an electrically pumped SPS and sources for entangled photon pairs such a sequential injection is potentially advantageous since it is typically difficult to achieve e.g. by Coulomb-blockade[13-15] but a key requirement for deterministic biexciton generation. The SAW-driven injection scheme requires a QD-system which is "strongly coupled" to the QW in which the charge conveyance occurs. This condition is met for most semiconductor heterostructures.

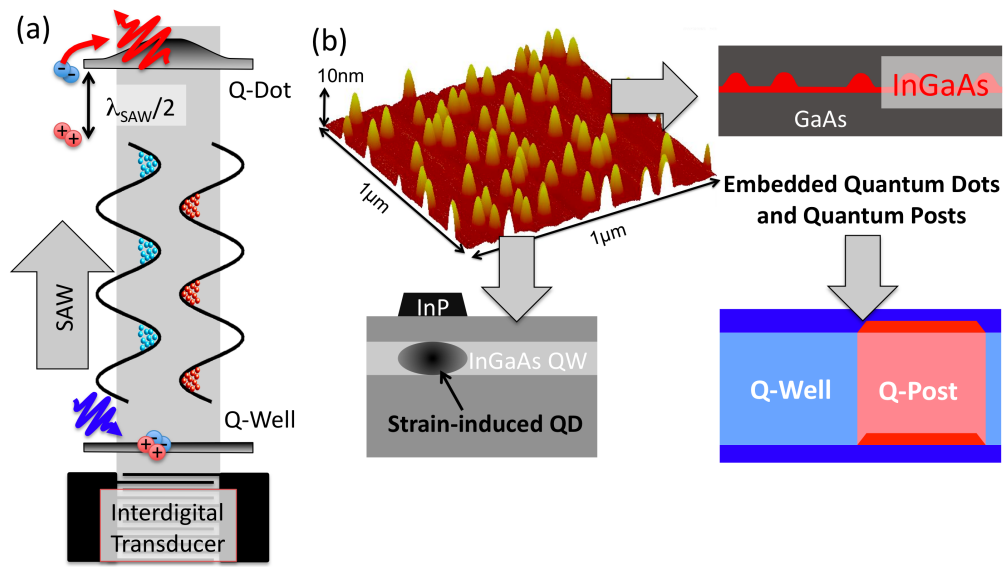

Figure 1 (color online): (a) Sketch of a SAW-driven SPS consisting of (bottom to top) an interdigital transducer (IDT) to excite a SAW, a generation area for electron hole pairs, a transport region where spatially separated electrons and holes are transported in the plane of a QW and finally a QD into which the transported carriers relax emitting a train of single photons. (b) Candidate QD-systems for an SAW-driven SPS based on self-assembled islands: (right) In(Ga)As QDs and QPs embedded in a GaAs matrix. (bottom left) Strain-induced QDs defined in a surface-near QW by an InP stressor on the sample surface.

The different QD systems explored in this work are presented in Fig. 1 (b). All three approaches are based on nanoscopic islands formed in a self-assembly process in the Stranski-Krastanov growth mode. An atomic force micrograph example of such InAs islands on a GaAs surface is shown in the upper left part of Fig. 1 (b). These islands can either create a zero-dimensional potential in a different heterostructure or form themselves an optically active QD.

In the first case [lower left part of Fig. 1 (b)], a strain-induced QD (SI-QD) is defined within a surface-near InGaAs/GaAs QW by the penetrating strain-field of an InP stressor island on the sample surface [19]. Since this type of dot is formed directly inside of a QW, the main requirement for SAW-driven injection is inherently fulfilled. Moreover, this system offers the unique possibility to post-growth isolation of individual QDs by selective removal of the InP stressors.

The second situation exists for self-assembled Stranski-Krastanov In(Ga)As QDs (SK-QDs, top right) and Quantum Posts (QPs, bottom right) [20,21] being embedded in a GaAs matrix. Here, the QD and QP themselves form the zero-dimensional confinement potential for both electrons and holes. In addition, both systems fulfill the second requirement of a coupled QW. In the case of the SK-QD the underlying wetting layer (WL) on which these islands form resembles a thin QW with clear two-dimensional properties [22]. The growth of self-assembled QPs is facilitated by a deposition sequence during which the height of a SK-QD is increased iteratively. During the formation of this columnar structure lateral In-segregation takes place, giving rise to the formation of a lateral QW-like matrix. Thus, both embedded QD-systems, SK-QDs and QPs, fulfill the two requirements for an SAW-driven SPS, namely zero-dimensional confinement and a coupled QW which can act as the transport channel.

## 3. STRAIN-INDUCED QUANTUM DOTS

### 3.1 Isolation of defined sub-ensembles and individual strain-induced Quantum Dots

The definition of SI-QDs in a QW by stressor islands on the sample surface provides the unique possibility to selectively remove a well-defined number of stressors and thus control the number and surface density of these dots from large ensembles down to individual, isolated SI-QDs.

Our sample consists of SI-QDs, which are formed by the strain field induced by InP islands in a 4 nm wide $In_{0.1}Ga_{0.9}As$/GaAs quantum well, located 5 nm below the sample surface. The ensemble of 60–80 nm diameter InP stressors initially has an areal density of $3\times10^9$ cm$^{-2}$ from which we select small sub-ensembles down to individual dots by selective removal of the surrounding islands using wet chemically using HCl and a mask defined by electron beam lithography [23]. This method allows for a precise control of the number of SI-QDs and we are able to perform spectroscopy on individual SI-QDs without using optical near-field or masking techniques. We study these samples by low-temperature ($T$ = 10 K) micro-photoluminescence ($\mu$-PL). Two spectra taken from a field of SI-QDs and in the etched region between are shown in Fig. 2 (e) as black and gray lines, respectively. While in the etched region only emission from the QW is detected at $\lambda_{QW}$ = 857 nm, the PL signal from the non-etched region is dominated by the SI-QD signals with ground state energies at $\lambda_{QD0}$ = 885 nm which is broadened due to the inhomogeneity of the stressor ensemble.

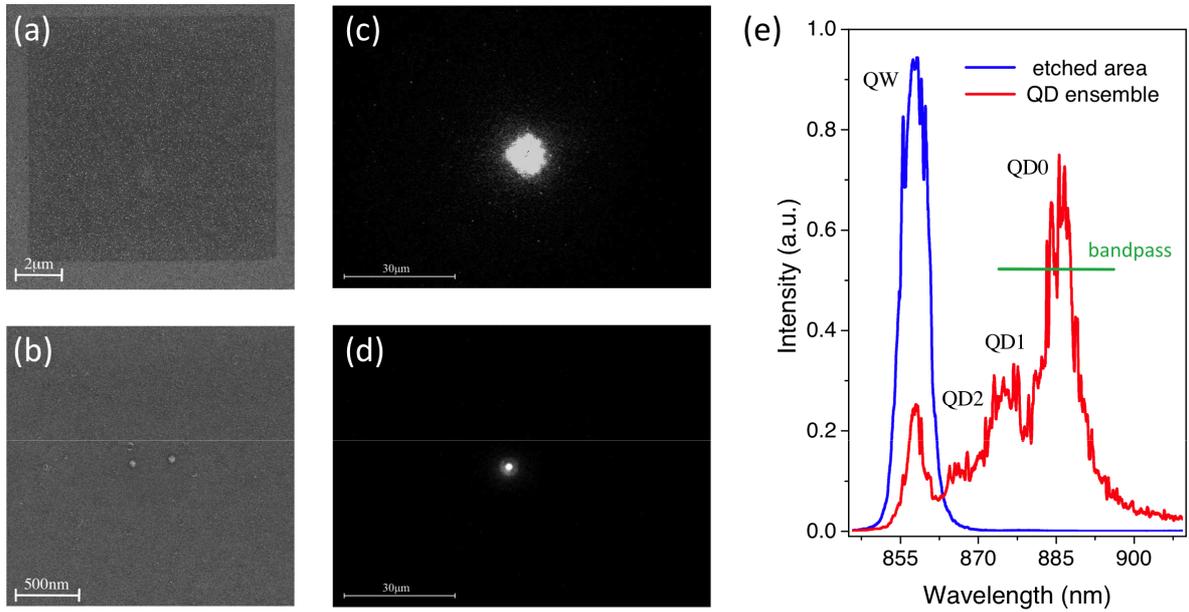

Figure 2: Isolation of SI-QDs – (a) and (b) SEM images of two sub ensembles containing approximately 3000 and exactly 2 SI-QDs. (c) and (d) represent direct PL-images of the two QD fields shown in (a) and (b) through a bandpass filter [horizontal bar in (e)] isolating SI-QD luminescence. (e) PL-spectra taken within a field containing QDs (black) and in an etched region (gray) where only signal of the QW is detected.

Emission of the SI-QDs is selected using a bandpass filter and directly imaged onto an intensified CCD camera. Here, the sample was excited using a 675 nm laser diode. SEM images of two isolated sub-ensembles with ~ 3000 [10 µm x 10 µm masked area] and exactly two QD [0.5 µm x 0.5 µm masked area] are shown in Fig. 2 (a) and (b), respectively. The direct PL-image of the larger square [Fig. 2 (c)] is clearly resolved and the square shape can be clearly identified. In the direct PL-image of the square containing exactly two SI-QDs their emission is clearly resolved. However, due to the small size of the emitting area, the image is diffraction limited. Along the route to further decrease the size of the sub-ensembles, we were finally able to isolate single SI-QDs which were investigated by time-integrated and time-resolved-PL spectroscopy and will be presented in the following section.

### 3.2 Cascaded exciton emission

After isolation of individual SI-QDs [SEM image in Fig. 3 (b)], we performed power dependent PL spectroscopy to assess the fundamental optical properties of the zero-dimensional system. An example of a typical excitation power series taken under pulsed excitation is shown in Fig 3 (a): At low excitation power ($P_0$ ~ 0.8 mW), only one emission line at 878 nm is observed which is attributed to the single exciton (1X=1e+1h) confined in the SI-QD. With increasing laser power, a second sharp peak appears at 879 nm, saturating at the same high excitation power (15 $P_0$) as the 1X exciton. This second peak is attributed to the biexciton (2X=2e+2h) with a binding energy of 1.5 meV. The broadening of 1X and 2X, respectively, is caused by additional charge carriers in the quantum dot during the s-shell decay. Our assignment is further supported by the intensity variation of the two peaks as a function of the excitation power: When plotted in double-logarithmic representation in Fig. 3 (b), the characteristic linear and quadratic dependence is observed for the 1X and 2X line, respectively. Furthermore, we want to note that the linewidths of the 1X and 2X peaks of 0.8 meV and 1.1 meV are not limited by the resolution of our spectrometer. Furthermore, both lines do not show any polarization anisotropy or direct evidence for a fine structure splitting [24]. This property makes this type of QDs potentially attractive as emitters of pairs of polarization entangled photons from the 2X-1X cascade.

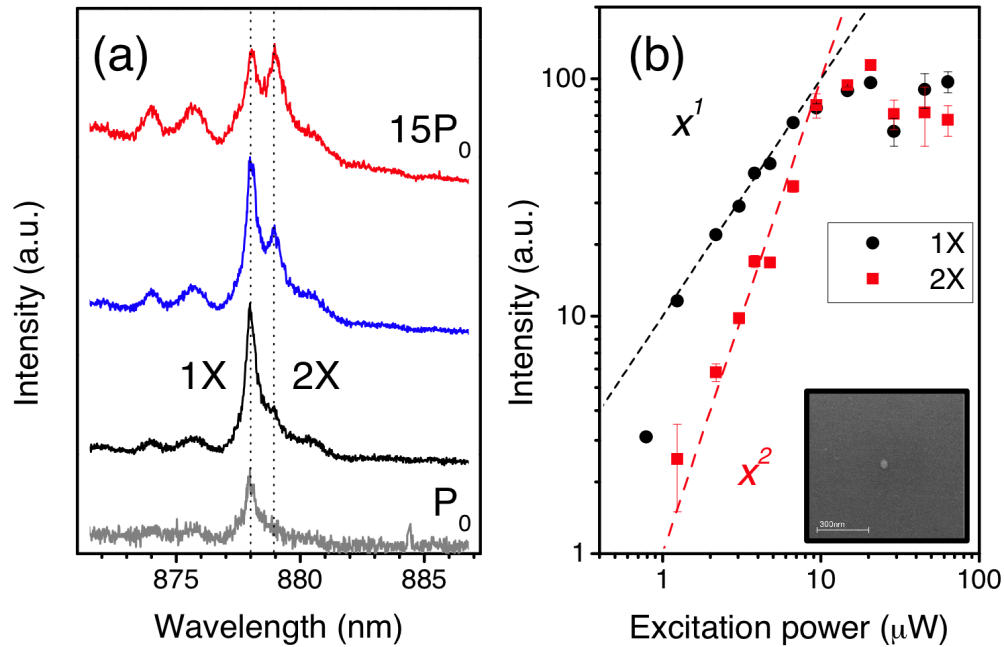

Figure 3: Single SI-QD spectroscopy – (a) Identification of 1X and 2X emissions from in the power dependent PL spectra. (b) Extracted line intensities as a function of excitation power.

In addition to the 1X and 2X lines shown in Fig. 3 (a), emission of higher order exciton complexes can be identified. In order to verify cascaded single photon emission we performed time-resolved PL spectroscopy on the three lowest exciton complexes, 1X, 2X and 3X. The 3X emission is detected ~ 15 nm blue-shifted compared to the 1X and 2X lines [23]. In these experiments, a single SI-QD was optically excited by a pulsed Ti:sapphire laser tuned to 818nm to resonantly excite the free exciton in the GaAs matrix. The emission of the different exciton complexes was spectrally isolated and detected by a Silicon single photon counting module. The overall time-resolution in these experiments was < 350 ps. We performed time-resolved PL spectroscopy on the 3X, 2X and 1X lines at three different excitation power levels (2 $\mu$W, 7 $\mu$W, 19 $\mu$W). The recorded time transients of the three transitions taken at the intermediate power level are plotted as solid lines in Fig. 4 (a) together with the instrument response function (dashed lines). When analyzing the exponential fits to the rising and falling parts we find e.g. that the decay time of the highest level (3X) $t_{fall,3X}$ = 0.9±0.35 ns agrees very well with the rising time $t_{rise,2X}$ = 1.1±0.35 ns of the next lower level (2X). The same behavior is observed with the decay of 2X and the rise of 1X with time constants of $t_{fall,2X}$ = 1.4±0.35 ns and $t_{rise,1X}$ = 1.5±0.35 ns. When considering a cascaded emission starting from 3X back to the crystal ground state (cgs) shown in Fig. 4 (b) this observation can be explained as follows: If the quantum dot is in the 3X state (3 electrons and 3 holes), there are two electrons and holes in the corresponding s-shells and one in the p-shells. After the decay of the p-shell e-h pair the dot is occupied by two electrons and two holes in the s-shell. At this stage recombination resulting in a photon with the energy of 2X can occur and, therefore, a state with three excitons in a QD delays the 2X transition as observed in the experiment by the correlation of the rise and fall times of these two signals. The same explanation applies to the 2X-1X signals. In addition, the observed onset of the 1X decay is delayed by the sum of the 3X and 2X fall times consistent with a cascaded emission of single photons [23,25].

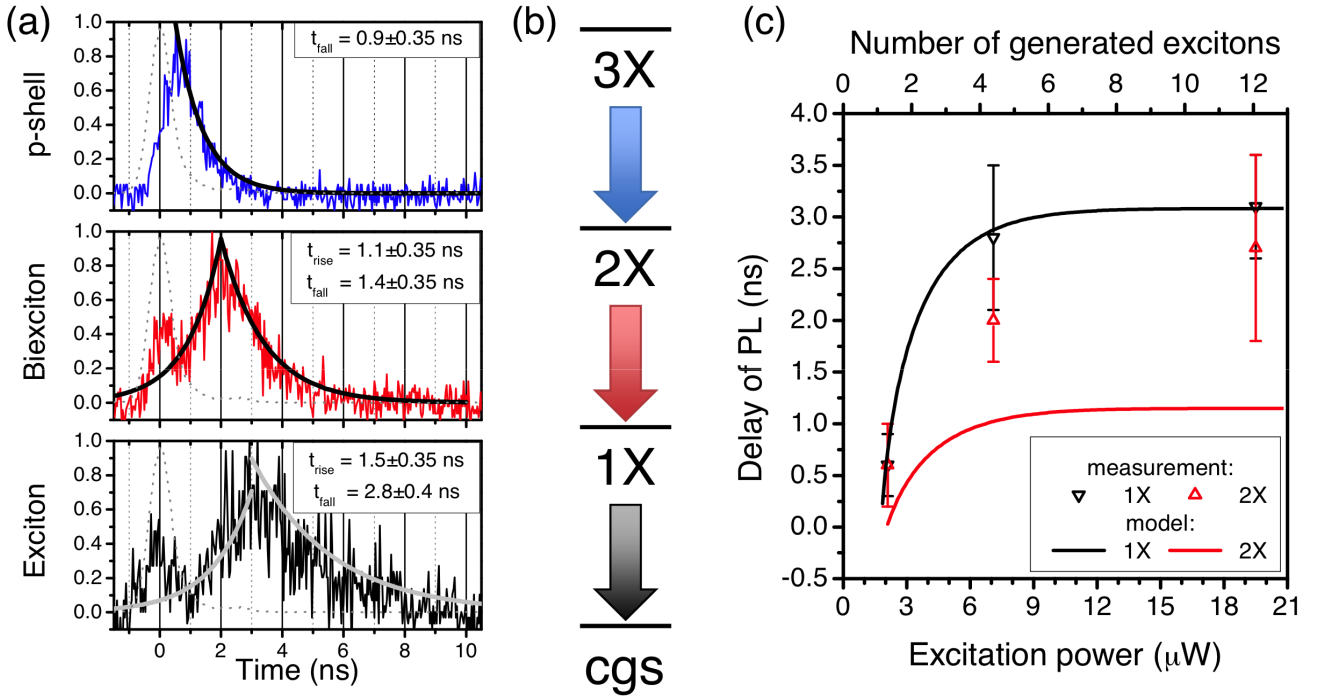

Figure 4: (a) Time-resolved PL-signals of the 3X, 2X and 1X emissions (solid lines) and the instrument response function (dashed lines). (b) Three-level cascade used to model the experimental data. (c) Calculated (lines) and measured delay of the PL signals of the 1X and 2X emissions.

Our findings represent a clear indication for cascaded emission of photons from a single SI-QD: The decay time of a higher level is correlated to the rising time of the next lower level and the onset of the emission from lower levels is delayed with increasing excitation power due to population of higher levels. This observation can be reproduced well by a rate equation model based on a three level cascade. This property in combination with the absence of fine structure splitting and polarization anisotropy makes SI-QDs potential candidates as emitters of polarization entangled photon pairs.

To further support these observations, we use a rate equation model to the system to reproduce the observed data. We restrict this model to the first three levels [Fig. 4 (b)] and neglect relaxation effects since the underlying time scales are much shorter than radiative decay times. In addition, we assume Poisson capture statistics of the photogenerated carriers into the QD levels. This is equivalent to the fact that the capture probability is independent of the population of the QD. We obtain the following rate equation

$$\frac{dn_i}{dt} = G_i - \frac{n_i}{t_i} \quad (1),$$

with $n_i$ being the occupation probabilities, $G_i$ the corresponding optical generation rates of charge carriers and $\tau_i$ the decay times of levels $i=1, 2, 3$. The generation rate of the highest level of the cascade is determined by the Poisson distribution with $g$ being the number of excitons generated by one laser pulse. In contrast, the occupation of the lower levels of the cascade has an additional contribution given by the decay rate of the next upper level:

$$G_i = \frac{n_{i+1}}{t_{i+1}} + e^{-i\frac{g^i}{i!}} \quad (2),$$

In order to solve equations (1) and (2) analytically, the parameters $\tau_i$ have to be determined experimentally. This was achieved by setting the excitation power to a level at which state $i$ is the highest occupied level. Thus, we ensure to measure the intrinsic decay constant of level $i$ since no effects from decays of higher levels are expected. With this set of input parameters we can reproduce the experimentally observed saturation behavior of the different emission lines with increasing excitation power. Furthermore, we can calculate the delayed onset of the different signals. Fig. 4 (c) shows the results of the model (lines) in comparison to the experimental data (points) for 1X and 2X. The calculated and experimental values for 1X are consistent and in good agreement. The deviation between our model and the experimental data for 2X can be explained by our restriction to a three level cascade.

### 3.3 SAW-driven carrier injection into a small ensemble of strain-induced Quantum Dots

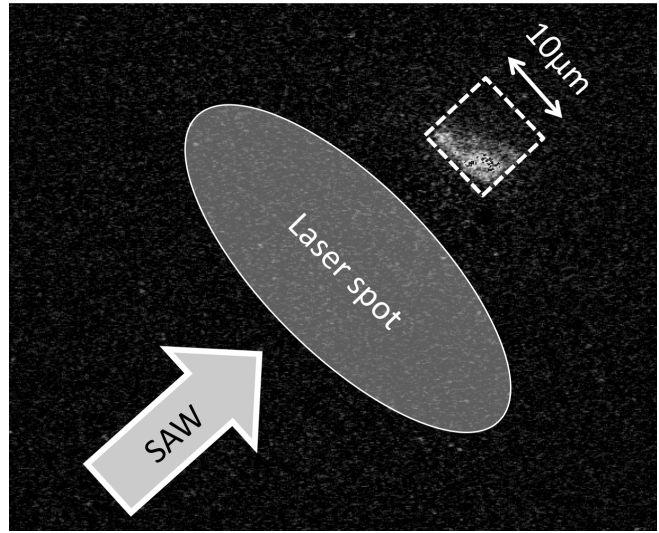

Figure 5: SAW-pumping of a small ensemble of SI-QDs. The total area of the QD-field, the position of the laser spot and the propagation direction of the SAW are marked accordingly.

In the experiments presented so far, the SI-QD were located directly inside the laser spot. In a next step, the SAW-driven injection scheme for the charge carriers into the QDs was realized. Here, the laser spot was displaced from the SI-QD sub-ensemble towards the position of an IDT exciting a SAW. Thus, the photogenerated electrons and holes are separated by λ_SAW/2 in real space and are transported with the propagating SAW in the plane of the QW. As the SAW arrives at the position of the SI-QD sub-ensemble, the transported charge carriers relax into the QD levels where after one SAW period excitons are formed. These excitons recombine radiatively and emit a train of single photons. In Fig. 5 (a) a direct PL-image of the SI-QD emission is shown and the position of the laser spot, the propagation direction of the SAW and the position of the QD field are marked. Clearly, emission is only detected from the QD field due to carrier transport and injection by the SAW [26]. In contrast to the directly excited PL image of the same field [Fig. 2 (c)], emission is only detected from areas facing towards the SAW and laser spot. This corresponds to approximately one quarter of the total area of the isolated sub-ensemble of SI-QDs containing about 750 SI-QDs which are opto-acoustically pumped.

## 4. SELF-ASSEMBLED QUANTUM DOTS AND QUANTUM POSTS

In the last part of this report, we describe the process flow to fabricate samples for SAW mediated carrier injection into isolated buried nanostructures such as QDs and QPs. This procedure can be readily applied for both systems QD and QP. In the following, we restrict the discussion to QPs and show that using this method we were able to transport electrons and holes using a SAW from the point of their generation to the position of isolated QPs giving rise to optical emission.

### 4.1 Alignment of IDTs on low surface density ensembles for SAW-driven carrier injection into single Quantum Posts

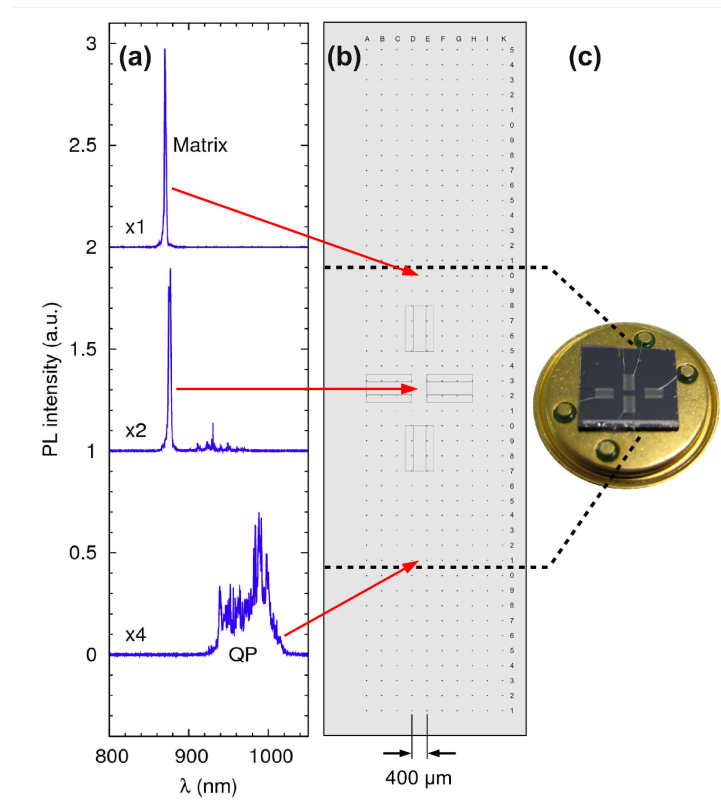

Figure 6: Alignment of IDTs with respect to a QP surface density gradient. (a) and (b) PL spectra taken with respect to a lithographically defined grid. (c) Photograph of the mounted sample.

In order to perform spectroscopy on single QPs in a µ-PL setup, low surface densities of these quantum structures are required. The density should be low enough to isolate a single QP but not too low in order to locate a QP in an adequate time. A convenient approach to fabricate material with a suitable surface density is to realize a gradient along one axis of the substrate by stopping the wafer rotation during deposition of the QP layers [27,28]. We start with a substrate with high QP density on the lower side of the wafer and no QPs on the upper side. To find the right density in-between we first cleave an approx. 20 mm long and 5 mm wide piece out of the wafer on which we expect the suitable region. Using electron beam lithography, crosshair markers are processed on this wafer piece. These markers are positioned on a grid with a sufficiently wide spacing to position IDTs in a later step. An example of such a layout is shown in Fig. 6 (b) for which the distance between two single markers is 400 µm. The entire piece is scanned from the bottom to the top and PL spectra are recorded for every point on the grid. Fig. 6 (a) shows three example spectra in the high surface density area (lower spectrum), the low surface density area (middle spectrum) and a region without QP (upper spectrum), respectively. Arrows in Fig. 6 indicate the positions at which these spectra were taken. As the sample is traversed, the characteristic exchange between the emission of the QPs and the surrounding matrix is observed. In the center part, a few individual sharp emission lines are observed which originate from isolated QPs which can be studied individually.

After this characterization, we isolate the region of low QP density from the wafer piece and fabricate IDTs in the third step. The outlines of four IDTs forming two delay lines are shown in the Fig. 6 (b). The IDTs were fabricated aligned to the grid used for determining the low surface density region. The vertical delay line is longer than the horizontal one to cover a wider range of QP density. Finally, the sample is mounted on a sample holder and the contact pads of the IDTs are wire bonded to the contact pins of the standard transistor sample holder. A photograph of the final sample is presented in Fig. 6 (c).

## 4.2 SAW-driven remote pumping of individual Quantum Posts

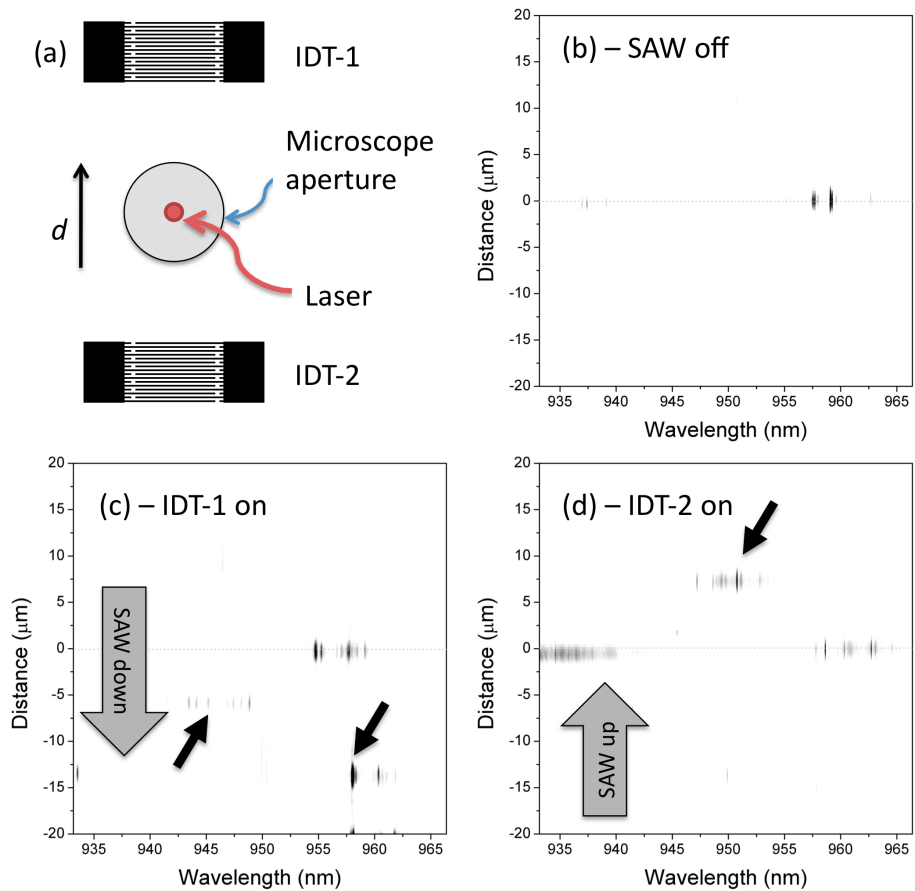

Figure 7: SAW pumping of individual QPs: (a) Sample layout, CCD image (b) without SAW, (c) IDT-1 turned on and (d) IDT-2 turned on

To demonstrate the SAW-driven pump scheme described in Section 2, we photo-excite carriers in a limited area and collect emission over a significantly larger area. In this scheme, depicted schematically in Fig. 7 (a), we focus the excitation laser ($\lambda$ = 661 nm, P = 300 nW) to a diffraction limited spot using a 50x microscope objective. The same objective collects the emitted PL light over its entire field of view with a diameter of 40 µm. We scan the sample to locate an individual QP position within the vertical delay line. The upper and lower IDT of this delay line are labeled IDT-1 and IDT-2 in Fig. 7 (a) and have resonance frequency of 193 MHz which corresponds to a $\lambda_{SAW} = 15$ μm.

The emission from the sample is dispersed in an imaging monochromator and detected by a cooled Silicon charge coupled device (CCD). The columns of the chip encode the spectral information whereas the rows contain spatial information. Thus, we align the propagation direction of SAWs excited by IDT-1 and IDT-2 with the rows of our CCD to recover the position of emitters along this direction $d$ indicated by the arrow in Fig. 7 (a). Thus, by reading out the entire CCD chip, spatial information along this direction of the sample is preserved. An example of such a one-dimensional PL-image for which the PL-intensity is encoded in grayscale is shown in Fig. 7 (b) for the case without any SAW excited. Clearly, only emission lines from a single QP are detected in a narrow region at the center of the CCD chip. The emitting QP is located inside the excitation laser spot and marks the center of the CCD chip. When a SAW is excited by the upper IDT-1, the spatial position remains unaffected and in addition, two new sets of lines appear in the lower part of the CCD. These signals arise from exciton recombination in QPs located between the laser spot and IDT-2. Since IDT-1 is operated, the SAW will transport carriers from the position of the laser spot i.e. towards IDT-2. These carriers are captured into QPs along the SAW propagation direction where they radiatively recombine. Moreover, we can determine the distance of these remotely pumped QPs to 7 μm and 14 μm, respectively. This finding clearly shows that the matrix quantum well surrounding the QPs provides a suitable transport channel for SAW-driven remote carrier injection on distances exceeding several tens of microns. To further confirm remote carrier injection, we reversed the SAW propagation direction by generating SAWs by IDT-2 instead of IDT-1. The obtained PL image is shown in Fig. 7 (d) and, as expected, no PL is detected from the lower part of the sample and the signal of a remotely pumped QP appears in the upper part.

These observations resemble a proof of principle of SAW-driven remote carrier injection into individual buried, self-assembled nanostructures and clearly show that this scheme can be applied for individual QPs and QDs.

## 5. SUMMARY

In summary, we have reported on recent progress towards acoustically pumped and triggered single photon emitters based on self-assembled quantum dot nanostructures. Our experiments on SI-QDs demonstrate single photon emission from the 2X-1X cascade and acoustically driven injection of carriers in a sub-ensemble containing ~750 QDs. We showed that this scheme can be readily extended to individual self-assembled QDs and QPs for which high quality single photon emission has been demonstrated [1,29]. Our scheme can also be applied to other types of QDs e.g. formed by thickness fluctuations in a quantum well as reported recently [30].

## 6. ACKOWLEDGEMENTS

This work was financially supported by the German Government as part of the Cluster of Excellence "Nanosystems Initiative Munich" (NIM).